\documentclass{aastex62}
\submitjournal{The Astrophysical Journal Letters}

\shorttitle{CIR Acceleration near 1.5 AU}
\shortauthors{Thampi et al.}

\begin{document}

\title{\textsc{Acceleration of energetic ions in corotating interaction region near 1.5 AU:\\ 
Evidence from MAVEN}}

\correspondingauthor{Smitha V. Thampi}
\email{smitha\_vt@vssc.gov.in}

\author[0000-0002-0786-7307]{Smitha V. Thampi}

\author[0000-0002-3324-156X]{C. Krishnaprasad}
\author[0000-0002-0979-0833]{P. R. Shreedevi}
\altaffiliation{Now at School of Space and Environment, Beihang University, Beijing, China}
\author{Tarun Kumar Pant}
\affiliation{Space Physics Laboratory, Vikram Sarabhai Space Centre, Thiruvananthapuram, India}
\author[0000-0003-1693-453X]{Anil Bhardwaj}
\affiliation{Physical Research Laboratory, Ahmedabad, India}



\begin{abstract}
The dearth of observations between 1 AU and 3 AU limits our understanding of energetic particle 
acceleration processes in interplanetary space.
We present the first-of-their-kind observations of the energetic particle acceleration in 
a Corotating Interaction Region (CIR) using data 
from two vantage points, 1 AU (near Earth) and 1.5 AU (near Mars). 
The CIR event  of June 2015  was observed by the 
particle detectors aboard  the Advanced Composition Explorer (ACE) satellite as well as the SEP 
(Solar Energetic Particle) instrument aboard the Mars Atmosphere and Volatile EvolutioN (MAVEN) 
spacecraft situated near 1.5 AU. We find that a CIR shock can accelerate a significant 
number of particles even at 1.5 AU. During this event the acceleration by the shocks 
associated with the CIR could cause an enhancement of around two orders of magnitude in the SEP  
energetic ion fluxes in the $\sim$500 keV to 2 MeV range when the observations near 1 and 1.5 AU are 
compared. To demonstrate the differences between SEP acceleration in CIR and other impulsive 
events, we show the energetic ion flux observations during an intense CME period in March 
2015, in which case the enhanced SEP fluxes are seen even at 1 AU. 
These observations provide evidence that CIR shock can accelerate particles in the region between 
Earth and Mars, that is, only within the short heliocentric distance of ~0.5 AU, in interplanetary 
space.
\end{abstract}



\section{Introduction} \label{sec:intro}

The existence of the continuous and supersonic solar wind predicted by \citet{Parker1958} is a 
consequence of the supersonic expansion of the solar corona. Fast solar wind 
streams originate from coronal holes, and as they interact with the slow solar wind in front of it, 
Corotating Interaction Regions (CIRs) are formed. The shocks associated with CIRs can accelerate 
energetic ions far beyond 1 AU, and these particles stream into the inner heliosphere 
\citep{VanHollebeke1978, Mason1999}. Unlike in the case of Coronal Mass Ejections (CMEs), the 
probability of CIR shocks forming within 1 AU is only $\sim$30\% \citep{Jian2006}, and whenever a 
solar energetic particle (SEP) event is observed at 1 AU, particle flux enhancements associated with 
these CIRs at larger heliocentric distances show a higher particle flux intensity \citep{Lario2000b}. The physical processes leading to the 
acceleration of these solar energetic particles 
are described in detail by \citet{Lee1982} and \citet{Desai2016}. The CIR associated energetic 
particles in the inner heliosphere were examined using observations by various spacecraft from 
1 AU to $\sim$9 AU \citep{VanHollebeke1978}. However, most of the vantage points for these 
observations, including Ulysses were beyond 3 AU, except for two   observations of Pioneer 11 at 
$\sim$1.5 AU, of which one of them was ambiguous \citep{VanHollebeke1978}, and a few  observations 
in mid\textendash 1992 and late 1994 when Ulysses moved from $\sim$5 AU near the ecliptic to a 
latitude of 80$^\circ$ S at $\sim$2.5 AU \citep{Richardson2004}. Apart from these isolated 
observations, most of the data points were from 1 AU \citep{Burlaga1985, Jian2006} and  
between 3 and 5 
AU \citep{VanHollebeke1978,  Burlaga1984b, Lario2000b,Crooker1999}.  \citet{Morgan2010} while studying the radar absorption events 
at Mars caused by 
energetic particles impinging on the Martian ionosphere mentioned that the forward and reverse 
shocks associated with a CIR evolve between 0.5 and 2 AU \citep{Crooker1999} and therefore the  
orbit of Mars, between 1.4 and 1.7  AU, is a key region to observe the acceleration of these 
energetic ions.\\

The Mars Atmosphere and Volatile EvolutioN (MAVEN) spacecraft was launched in November 2013 primarily to understand the atmospheric 
escape from Mars \citep{Jakosky2015a}. MAVEN has a unique suite of instruments 
to measure the solar wind velocity, solar wind particle density, interplanetary magnetic field 
(IMF), 
solar wind electron pitch angle distributions, solar EUV irradiance, and solar energetic electron 
and ion fluxes along with Martian ionosphere-thermosphere measurements. The  SEP
flux measurements from MAVEN  provide a new vantage point near $\sim$1.5 AU to study the 
acceleration of energetic particles  by CIR and CME driven shocks.\\


In this study, we compare the SEP fluxes during the CIR  event of June 2015 and CME events of March 
2015 as observed near Earth (1 AU) and near Mars ($\sim$1.5 AU) to understand the acceleration of 
the energetic particles beyond 1 AU, especially for the CIR driven event. The uniqueness of this 
study is that for the first time, we have a vantage point to study the energetic particle fluxes 
near 1.5 AU associated with CIRs. The CME events are shown to demonstrate the differences between 
the SEP fluxes in CIRs and other impulsive events at these vantage points. The direct measurements 
presented here, from these two key regions in space would be useful to understand the evolution of 
energetic particles in CIR and the changes in their spectral characteristics, within short 
heliocentric distances.\\

\section{Data}\label{sec:data}

The energetic particle measurements at 1 AU for the selected events are from the Electron Proton 
Alpha Monitor (EPAM) sensor onboard Advanced Composition Explorer (ACE) satellite. The EPAM  
provides particle fluxes in the energy range from 47 keV to 4.75 MeV, in 8 channels. We have used 
ion fluxes measured in 4 channels in the energy ranges, 0.31-0.58 MeV, 0.58-1.05 MeV,  1.05-1.89 
MeV, and 1.89-4.75 MeV by the Low Energy Magnetic Spectrometer of the EPAM sensor. These data are 
obtained from the ACE data center (\url{http://www.srl.caltech.edu/ACE/ASC/level2/}). The solar wind 
velocity,  the IMF near 1 AU and the Geostationary Operational Environmental Satellite (GOES) 
energetic particle flux variation ($>$ 10 MeV and $>$ 30 MeV) are obtained from the NASA Space 
Physics Data Facility (SPDF) OMNIWeb data center (\url{https://omniweb.gsfc.nasa.gov/}).\\

The solar wind velocity and IMF values at 1.5 AU are obtained from the Solar Wind Ion Analyzer 
(SWIA) and Magnetometer (MAG) instruments aboard MAVEN spacecraft. The SWIA and MAG data are used to 
compute the upstream solar wind parameters following the method by \citet{Halekas2016}. The SEP 
fluxes are obtained from the Solar Energetic Particle (SEP) instrument aboard MAVEN. This instrument 
consists of two identical sensors, SEP 1 and SEP 2, each consisting of a pair of double\textendash 
ended solid\textendash state telescopes  to measure 20 keV\textendash 1 MeV electrons and 30 
keV\textendash 6 MeV ions in four orthogonal view directions \citep{Larson2015}. The data used in 
this study  are the ion data  in the form of energy fluxes measured by the SEP 1 sensor in the 1F 
direction  that typically views the Parker spiral direction \citep{Larson2015}. The gaps in the SEP 
data are when the instrument attenuators close at altitudes below 500 km \citep{Lee2017}. These 
datasets from the MAVEN instruments are downloaded from the the Planetary Data System 
(\url{https://pds.nasa.gov/}). The level 2, version 01, revision 01 (V01\_R01) data of SWIA, level 
2, version 01, revision 01/02 (V01\_R01/R02) data of MAG, and level 2, version 04, revision 02 
(V04\_R02) data of SEP are used for the analysis. The Solar Dynamics Observatory (SDO) images 
are taken from \url{https://solarmonitor.org/}. The ENLIL simulations during June and 
March 2015 
are taken from ENLIL Solar Wind Prediction (\url{http://helioweather.net/}, 
\url{https://iswa.ccmc.gsfc.nasa.gov/}).\\ 

\begin{figure}[h]
\includegraphics[width=1.0\linewidth]{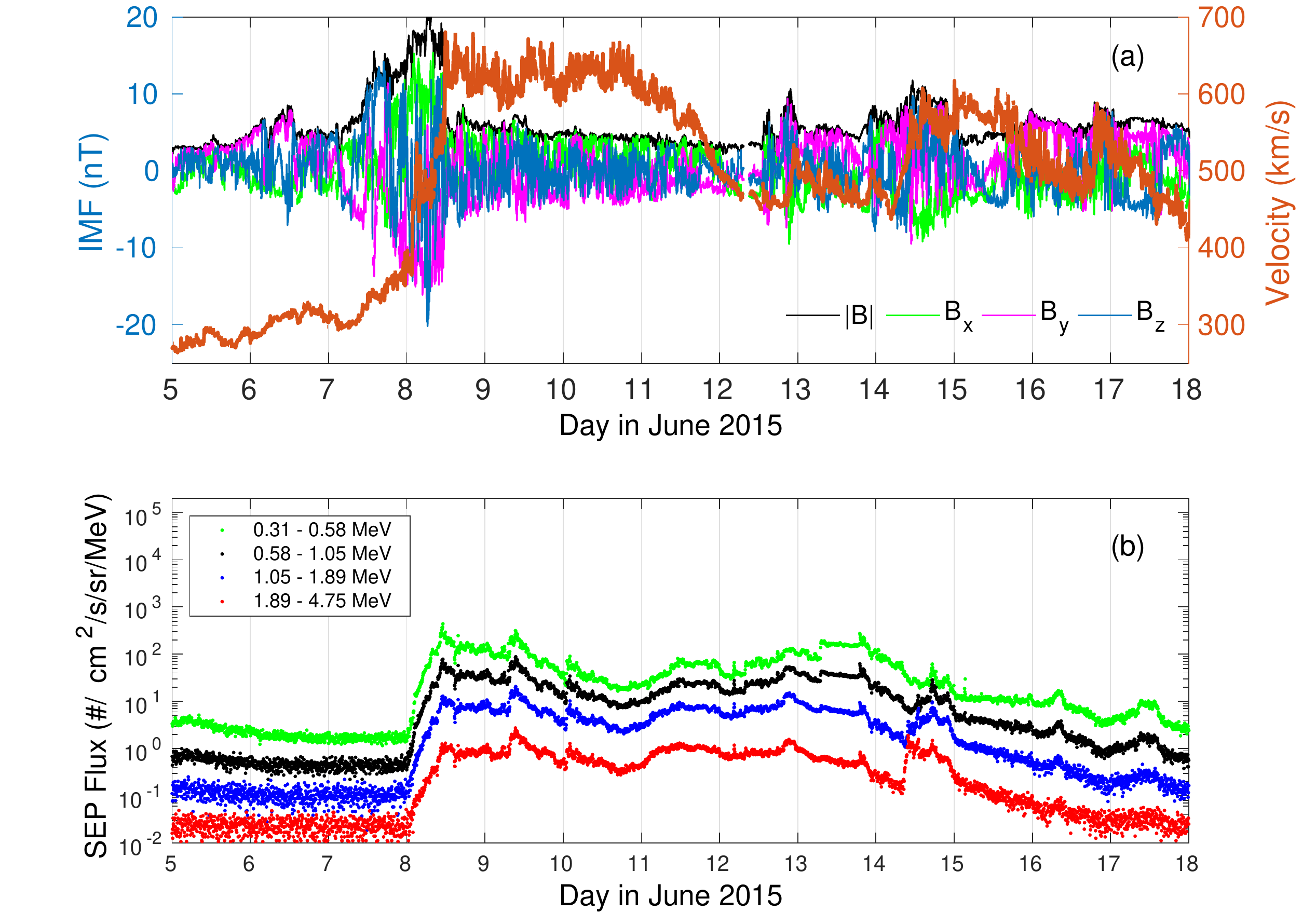}\\
\caption{(a) The variation of IMF (left Y-axis) and solar wind velocity (right Y-axis) for 
the 
period 5\textemdash17 June 2015 observed by ACE at 1 AU. (b) The SEP particle fluxes at different 
energy ranges as observed by the EPAM sensor onboard ACE satellite 
for the period 5\textemdash17 June 2015.}
\end{figure}

\begin{figure}[h]
\includegraphics[width=1.0\linewidth]{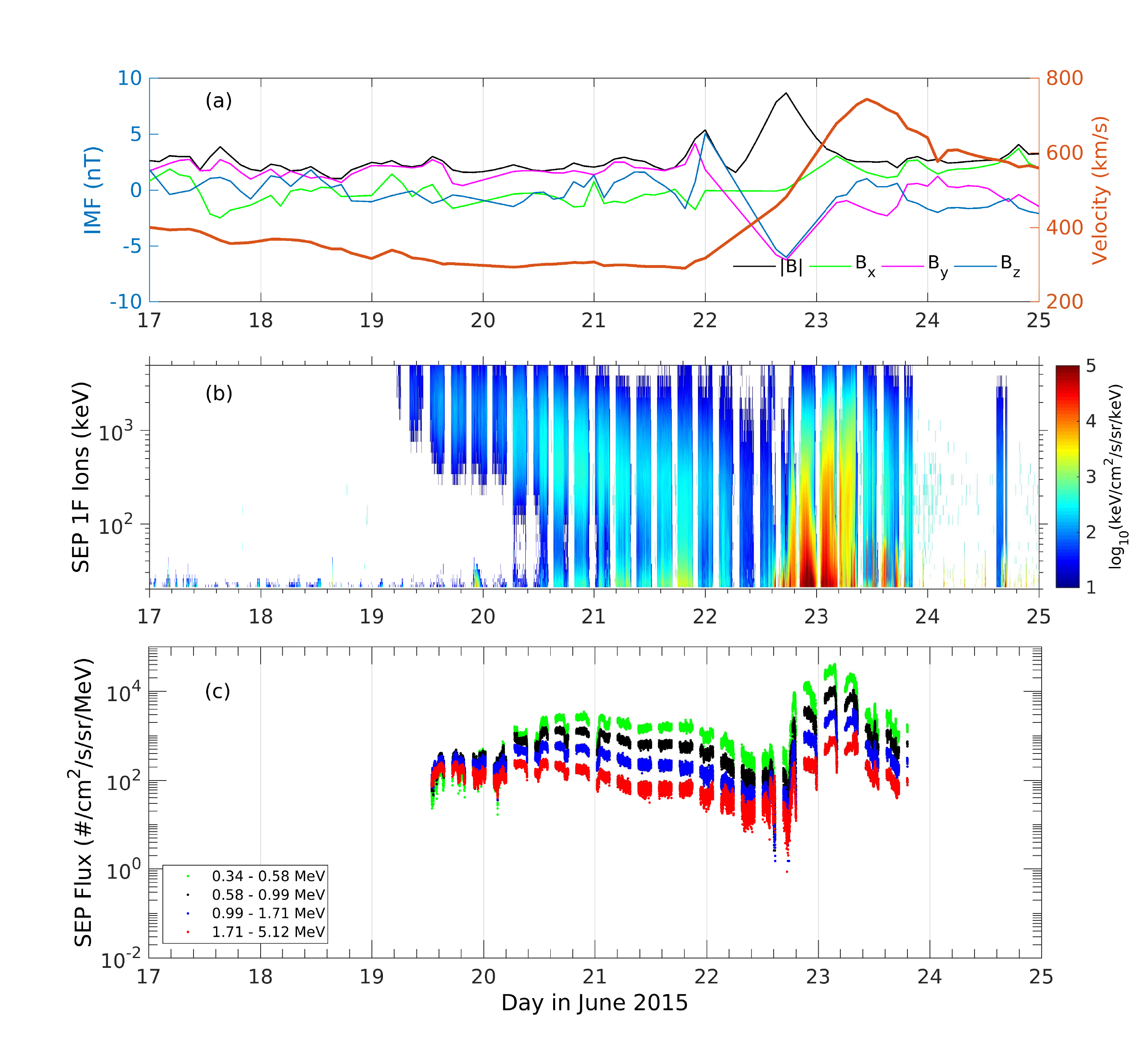}\\
\caption{(a) The variation of IMF  (left Y-axis) and solar wind velocity 
(right Y-axis) for the period 17\textemdash24 June 2015. (b)The energy spectrum of ions in the 30 
KeV to $\sim$6 MeV range observed by the MAVEN SEP instrument for the 
period 17\textemdash24 June 2015. The Y-axis shows the energy range and the color axis shows the 
logarithmic particle flux. The initial period shown in white had very low fluxes ($<$10) whereas the 
gaps on 24 June 2015 corresponds to absence of good quality data. (c) The integrated SEP particle 
fluxes in different energy ranges (comparable to ACE measurements) estimated from the MAVEN SEP 
data.}
\end{figure}

\begin{figure}[h]
\includegraphics[width=1.0\linewidth]{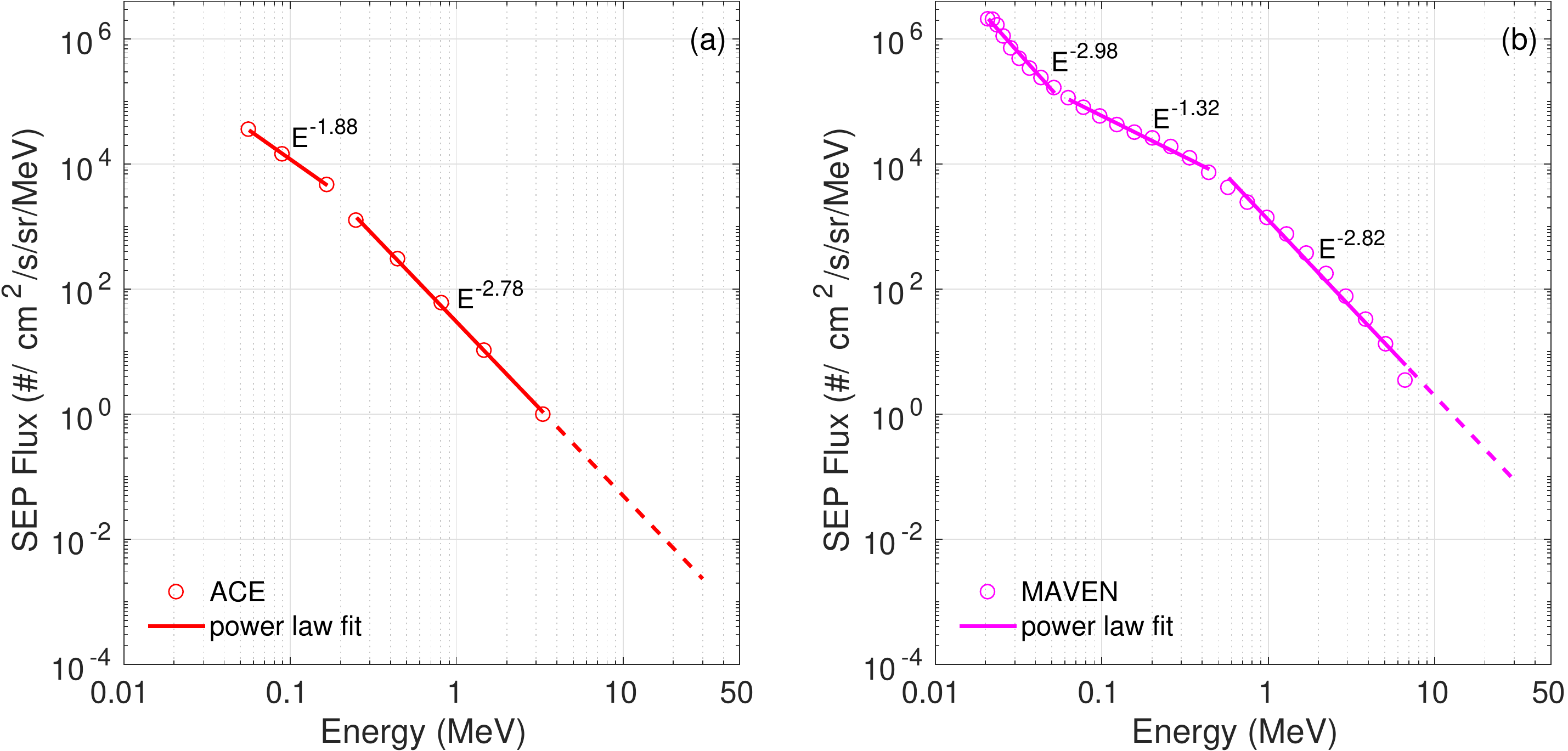}\\
\caption{ (a) The energy spectrum during peak of ion flux enhancement in the CIR event on 8 June 
2015 detected by ACE at $\sim$1 AU. The  power laws fitted with the observations are also shown. 
The observations by ACE are extrapolated to 30 MeV, to show that at both 10 MeV and at 30 MeV, the 
expected flux levels are below the detection limits \citep{Rodriguez2014} of GOES.  
(b) The energy spectrum obtained from MAVEN.}
\end{figure}

\begin{figure}[h]
\includegraphics[width=1.0\linewidth]{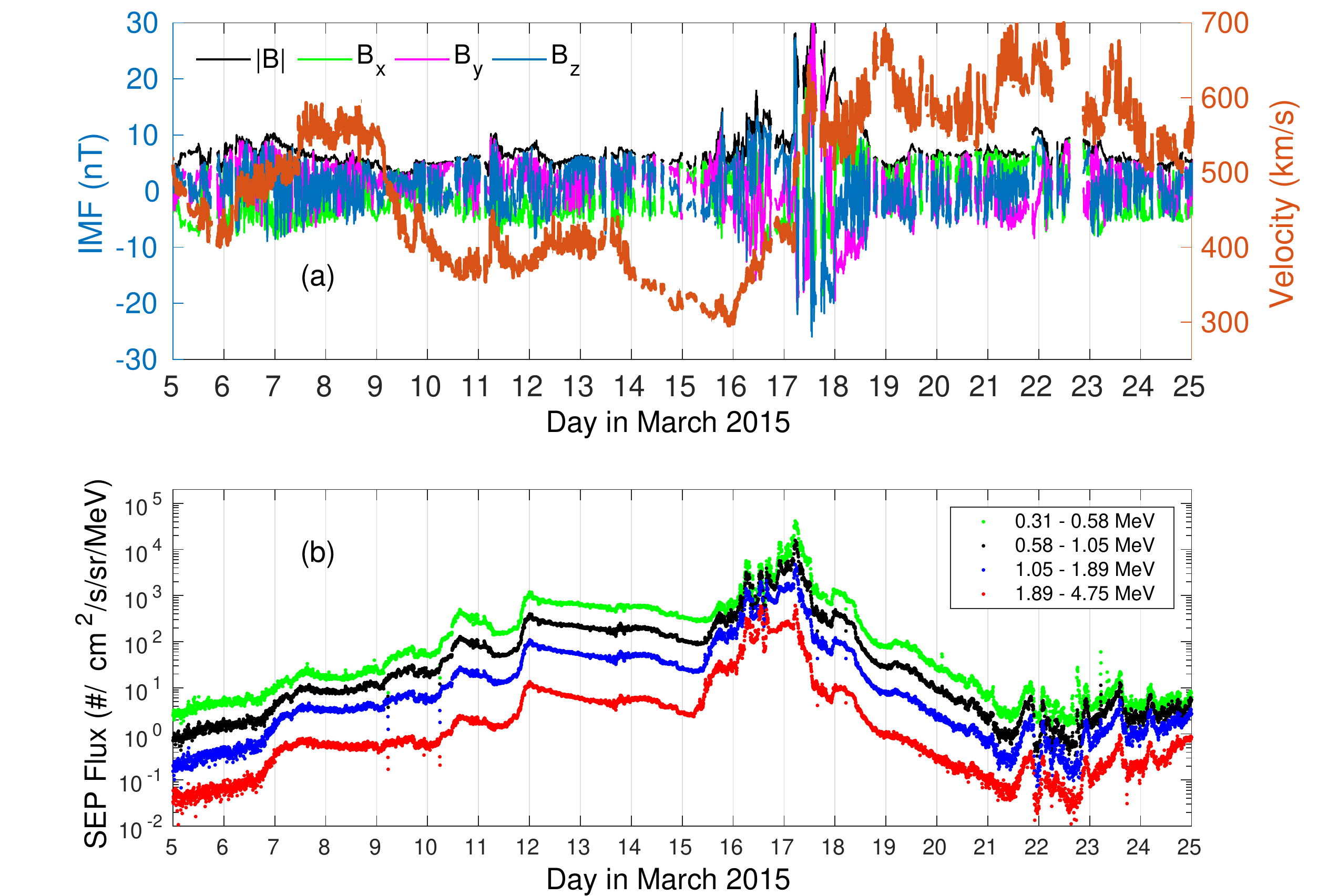}\\
\caption{Same as Figure 1, but for the period 5\textemdash24 March 2015.}
\end{figure}

\begin{figure}[h]
\includegraphics[width=1.0\linewidth]{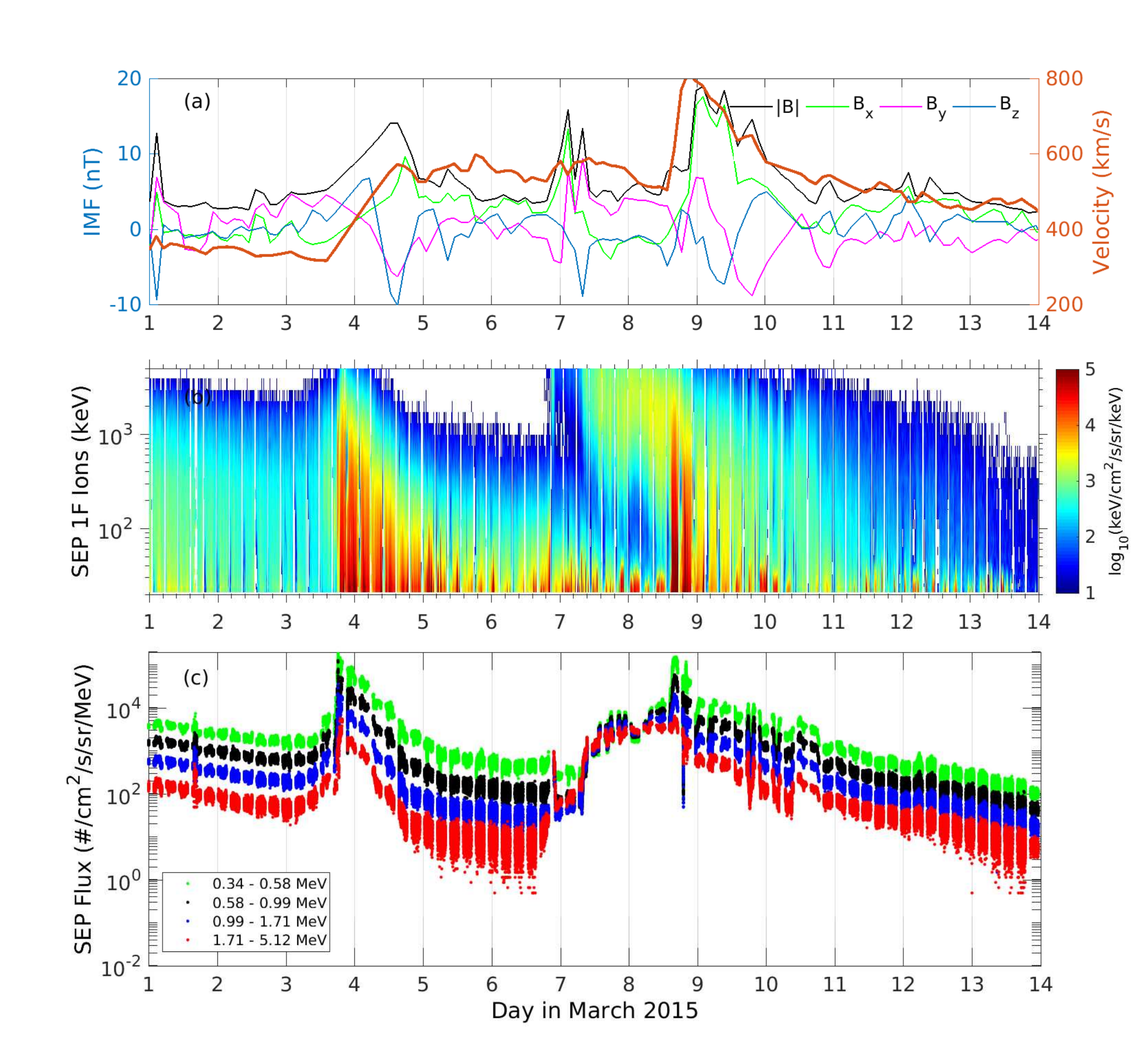}\\
\caption{Same as Figure 2, but for the period 1\textemdash13 March 2015. }
\end{figure}

\section{Overview of events}\label{sec:event}

A  comprehensive overview of the solar cycle 24 space weather conditions observed during the first 
$\sim$1.9 years of the MAVEN science mission had been provided by \citet{Lee2017}. The events 
described in this study are selected from \citet{Lee2017}. During June\textendash July 2015, a 
number of low\textendash latitude coronal holes appeared and generated recurrent high\textendash 
speed streams. The SDO/Atmospheric Imaging Assembly (AIA) composite images of the Earth-facing disk 
of the solar corona shows a persistent low-latitude coronal hole that remain more or less similar 
even after a 27 day solar rotation period. The images of the coronal hole are given in 
\citet{Lee2017}. The ENLIL predictions (see Appendix \ref{appendix}) reveal that the CIR 
hits Earth on 7 June 2015. The ENLIL snapshots further confirm that the same coronal hole 
with negative IMF polarity is rotated towards Mars and arrive at Mars on 22 June 2015.  Assuming to 
a first approximation that the rotation period for the coronal hole source is about 27.3 days, with 
Mars located at 1.56 AU and in solar conjunction with Earth the solar wind stream would arrive 
$\sim$15.5 days later (half a Carrington rotation $+$ $\sim$2.5 days to propagate from 1 AU to 1.56 
AU) after the observation at Earth on 7 June \citep{Lee2017}. The ENLIL simulations also 
show that at STEREO location the CIR stream would hit on 19 June 2015 (see \ref{appendix}). These 
simulations confirm that these streams are originated from the same coronal hole. Further, the study 
by \citet{Fisk1980} reports that typically, the high-speed stream is relatively steady in time, the 
CIR and the shocks associated with it are stationary in the frame corotating with the Sun. 
Therefore, the energetic ions from a single CIR event can be  seen for a period of 17 days and 
$\sim$225$^\circ$ in solar longitude. \citet{Reames1997} also find clear evidence of energetic ions 
from CIR shocks over a vast spatial region corresponding to far more than half a solar rotation.\\ %

End of February 2015 through early March 2015 was a period of intense space weather disturbances 
at Mars \citep{Lee2017}. 
Bright CME loops were observed in the 
southeast quadrant of the Solar
and Heliospheric Observatory Large Angle and Spectrometric Coronagraph (LASCO) 
C2 field of view on 6 March 2015, and peak enhancements in solar wind speed, density, dynamic 
pressure, and IMF were observed at Mars on 8 March 
2015 (\citet{Thampi2018} and references therein). The solar source, 
for these events, Active Region 12297 was  on the backside of the Sun (as viewed from Earth), which 
rotated over the east solar limb and produced a 
M8.9 class flare that was observed by both MAVEN and 
GOES on 7 March 2015 \citep{Lee2017}. As this active region rotated toward the direction of Earth, 
it produced a number of flares and CMEs including the 17 March 2015 superstorm at Earth, popularly known as the 
``St. Patrick's Day Storm''. 
ENLIL simulations confirm the CME arrivals and shows that these events at Earth and Mars 
are 
produced by different eruptions from the same active region (see \ref{appendix}). \\

\section{Observations}\label{sec:obs}

Figure 1a shows the variation of IMF and solar wind velocity as observed near 1 AU, during 
5\textendash 17 June 2015. The CIR reached the vicinity of Earth on 7 June 2015, marked by an 
enhancement in velocity. This enhancement in velocity continued to remain near the highest value  
almost for 4 days and then remained at a slightly lower value (which is still higher than the 
velocities observed during 5\textendash 7 June 2015) upto  17 June 2015. The Z\textendash component 
of the IMF showed largest fluctuations on 7 and 8 June, followed by fluctuations of smaller 
magnitude, a typical feature of a CIR driven storm \citep{Borovsky2006}. Figure 1b shows the 
variation of the energetic particle fluxes of energy range from $\sim$300 keV to $\sim$4.7 MeV as 
observed by the EPAM sensor onboard ACE satellite located near 1 AU. The signature of the CIR 
arrival is seen as $\sim$2 orders of magnitude increase in the SEP fluxes. The arrival of the CIR is 
preceded by a calm period with almost no SEP fluxes. This is another typical feature of 
CIR events \citep{Borovsky2006}. The flux of particles in the energy range $\sim$0.3 MeV to 
$\sim$0.5 MeV increased to $>$100 particles cm$^{-2}$s$^{-1}$sr$^{-1}$MeV$^{-1}$, whereas the 
particle flux with energies $>=$1 MeV remained $\le$ 10 particles 
cm$^{-2}$s$^{-1}$sr$^{-1}$MeV$^{-1}$.\\

Figure 2a shows the variation of IMF and solar wind velocity as observed 
by the MAG and SWIA instruments aboard MAVEN located at 1.56 AU. The total magnetic 
field increased to $\sim$9 nT coincident to the arrival of CIR. The solar wind velocity 
increased from the quiescent value of $\sim$300 km s$^{-1}$ to $\sim$700 km s$^{-1}$ and remained to 
be $>$ 600 
km s$^{-1}$ for the next day as well and remained high upto the end of June 
\citep{Lee2017}.\\

Figure 2b shows the differential energy flux spectrum of the solar energetic ions observed by the 
MAVEN SEP instrument. During the CIR event (on 22 and 23 June 2015), there is an enhancement in the 
flux of energetic particles. To see the actual number of accelerated energetic ions using the MAVEN 
SEP instrument and compare them with the observations at 1 AU by ACE, we estimated the integrated 
flux at energy bins comparable to that of EPAM/ACE observations (Figure 2c). The onset of disturbed 
fluxes from 8 June at Earth shown in Figure 1b (coinciding with the enhancements in velocity and 
magnetic field)  correspond to the enhancement seen from 22 to 23 June at Mars. The time differences 
seen in the SEP enhancement at these locations is consistent with the estimated arrival time based 
on the solar rotation period for the coronal hole source. It must be noted that the SEP fluxes 
before 19 June at Mars had very low values and the source of the slightly enhanced SEPs seen on 
19-20 June could be related to the streaming particles from field lines connected to a CME that 
erupted over the west solar limb, although the CME has not directly hit Mars \citep{Lee2017}. The 
SEP fluxes return to  lower values (below 100 particles cm$^{-2}$s$^{-1}$sr$^{-1}$MeV$^{-1}$) before 
the arrival of CIR on 22 June. The arrival of the CIR is marked by the enhancements in solar wind 
velocity and magnetic field, and coincident enhancement in SEP fluxes by more than 2 orders from 
the pre-CIR levels.  In comparison to the observations 
from 1 AU, fluxes enhanced in the similar energy ranges at 1.5 AU. The peak ion flux in the 
$\sim$1\textendash 2 MeV bin enhanced by $\sim$300 times at 1.5 AU. This indicates that  
acceleration took place beyond 1 AU, but within 1.5 AU. Similarly, $\sim$140 times enhancement is 
observed for the flux in the $\sim$600 keV\textendash 1 MeV energy bin. Therefore, we can infer that 
the acceleration by the shocks associated with CIR could cause an enhancement of around two orders 
of magnitude in the SEP fluxes when observed near 1.5 AU.\\

Figure 3a shows the energy spectrum corresponding to the peak of ion flux enhancement for the CIR 
event on 8 June 2015 detected by ACE. Double power laws are fitted, one with 
E$^{-1.88}$ for flux below 0.2 MeV and the second one with E$^{-2.78}$ for the flux between 0.2 to 
3 MeV. These spectral indices are similar to those reported by \citet{Bucik2009} for the May 2007 
events. To see whether we can expect higher energy particles at the energy ranges of GOES, 
the ACE spectrum is extrapolated to 30 MeV with the same power law. With this approximation 
itself, it can be seen that at both 10 MeV and at 30 MeV the expected flux levels are below the 
detection limits of GOES \citep{Rodriguez2014}. A double 
power law or a power law with exponential rollover at a few to tens of MeV has been reported in many 
events (e.g. \citet{Mason1999}).  If so,  the expected flux at the higher energy ranges will be 
still lower than 
what we estimated from a simple power law. Therefore, for this CIR event we do not expect that GOES 
would detect the SEP fluxes, and the observations confirm that GOES did not see any particles above its detection threshold.\\ 

Figure 3b shows the  energy spectrum during peak of ion flux enhancement in the CIR event on 23 
June 2015 detected by MAVEN, with power law fits. The first 
portion (below 60 keV) may have contributions from the oxygen pickup ions \citep{Larson2015}. The flux in the energy range 0.5\textendash3 MeV 
shows a spectral index of -2.82.   At both 
these 
locations, the observed energy spectra continued to extend as power laws right down to the lowest 
energy measured by the instruments, which corroborates with the results of \citet{Mason1997} and 
\citet{Chotoo2000}.\\

To demonstrate the difference between CIRs and other impulsive events, we 
show in Figure 4a (Figure 5a) the variation of IMF and solar wind velocity  
observed near 1 AU (1.5 AU), during 5\textendash 24 March 2015 (1\textemdash 13 March 2015). Figure 
4a shows that the IMF showed large fluctuations 
maximizing on 17 March. The solar wind velocity also showed concurrent 
enhancement and fluctuations afterwards. Figure 4b shows the SEP fluxes observed by the  ACE satellite for the same period. The fluxes showed 
enhancement in all 
the energy levels during the CME event. Compared with the CIR related energetic particle fluxes, 
these observations are more than 2 orders of magnitude higher in all the energy ranges observed by 
EPAM/ACE. This is the typical characteristic of CME driven shock acceleration of energetic particles 
and one of the major difference between CME and CIR driven storms at 1 AU. The GOES observations 
(not illustrated) also indicate the presence of particles in the $>$ 10 MeV energy range, that is, 
a peak flux of $\sim$7 cm$^{-2}$s$^{-1}$sr$^{-1}$ on 17 March.\\ 

Figure 5a shows the variation of IMF and solar wind velocity observed by  MAVEN for the period 
1\textemdash 13 March 2015 and Figure 5b shows the corresponding SEP energy flux spectrum.  The 
first  velocity enhancement occurred on 3 March and the highest velocity was observed on 8 March. 
IMF variations were observed during both the events with  higher magnitudes during the second event. 
The SEP spectrum showed flux enhancements with almost similar levels on both the occasions. MAVEN 
observed the CME shock only after the arrival of the highest energy SEP ions and these high energy 
($>$1 MeV) ions arrived first followed by the lower energy particles, during 7\textendash8 March. 
Figure 5c shows the integrated particle fluxes at Mars in the energy ranges comparable to EPAM/ACE. 
The flux levels showed increase at $\sim$1.5 AU at all the energy bins. In the energy range of 
$\sim$500 keV to 1 MeV, the enhancement is only  $\sim$5 times. \\


\section{Discussion and Conclusions}
Most of the previous studies concerning CIR driven SEP flux enhancements were 
confined to data from observations beyond 3 AU. The radial gradient of CIR associated energetic 
particle streams was measured by instruments on multiple probes to the inner heliosphere  
(Helios 1 and 2,  Mariner 10, Ulysses), and the outer solar system (Pioneer 10 and 11) with 
reference spacecraft at 1 AU such as IMP 7 and 8, ACE, WIND, and STEREO, and most of these 
measurements pertained to total ion flux in the 0.9\textendash 2.2 MeV energy range. The Ulysses observations 
at $\sim$1.4 AU, during its rapid latitude scan  phase  had shown the presence of reverse waves 
that had not yet steepened into shocks \citep{Gosling1995c}. Apart from this, there was one 
unambiguous observation of CIR induced acceleration of ions  near 1.5 AU, by Pioneer 11, which was 
total ion fluxes in the energy range 0.9\textendash 2.2 MeV \citep{VanHollebeke1978}. The 
enhancement in the fluxes reported here is almost an order of magnitude higher than that reported 
by \citet{VanHollebeke1978}. \\

The  observations presented here clearly illustrate the following aspects:\\

(a) During the CME events in March 2015, the SEP observations from both 1 and 1.5 AU show 
enhancement of high energy particle fluxes. The observations suggests that shock acceleration 
takes place at 1 AU itself during CME events \citep{Lario2000b}. The comparison facilitates to 
demonstrate the differences between SEP acceleration in CIRs and other impulsive events like 
CMEs.\\ 

(b) During the CIR event in June 2015, at 1 AU, GOES did not observe any particle flux above its 
detection threshold, and the particle flux $>=$1 MeV detected by ACE remained $\le$ 10 particles 
cm$^{-2}$s$^{-1}$sr$^{-1}$MeV$^{-1}$. Compared to this, the SEP instrument aboard MAVEN showed high 
SEP fluxes upto $\sim$3 MeV. Comparison shows that the acceleration by the shocks associated with  
CIR can produce an enhancement of around two orders of magnitude in the SEP fluxes when observed 
near 1.5 AU. Interestingly, the peak high energy particle fluxes at Mars are of almost similar 
levels during CME driven event indicating similar levels of acceleration during these events at 1.5 
AU.  In contrast, these flux levels are drastically different in the case of ACE observations at 1 
AU ($>$ one order of magnitude higher in the $\sim$1\textendash 2 MeV range during CME), consistent 
with the expectation that strong shock acceleration takes place inside 1 AU during CME events 
\citep{Li2005, Desai2008}.  The comparison of energy fluxes at different ranges may point toward the 
change of energy with distance as particles travel several AUs in the heliosphere under 
cross\textendash field diffusion \citep{Zhao2016}. These observations are unique because of the 
addition of a vantage point  at 1.5 AU, and as such observations were previously very sparse between 
1 and 3 AU.\\




\newpage
\appendix
\section{WSA-ENLIL+Cone model simulations}
\label{appendix}
We used the Wang-Sheeley-Arge (WSA)\textendash ENLIL+Cone model to numerically simulate 
the interplanetary solar wind plasma and magnetic field conditions and provide a global 
heliospheric 
context for the solar events discussed. In this heliospheric model, 
the solar coronal model WSA is coupled with the three-dimensional magnetohydrodynamic numerical 
model ENLIL, which is combined with the Cone model \citep{Odstrcil2003, Mays2015}.  

\begin{figure}[h]
\centering
\includegraphics[width=0.7\linewidth]{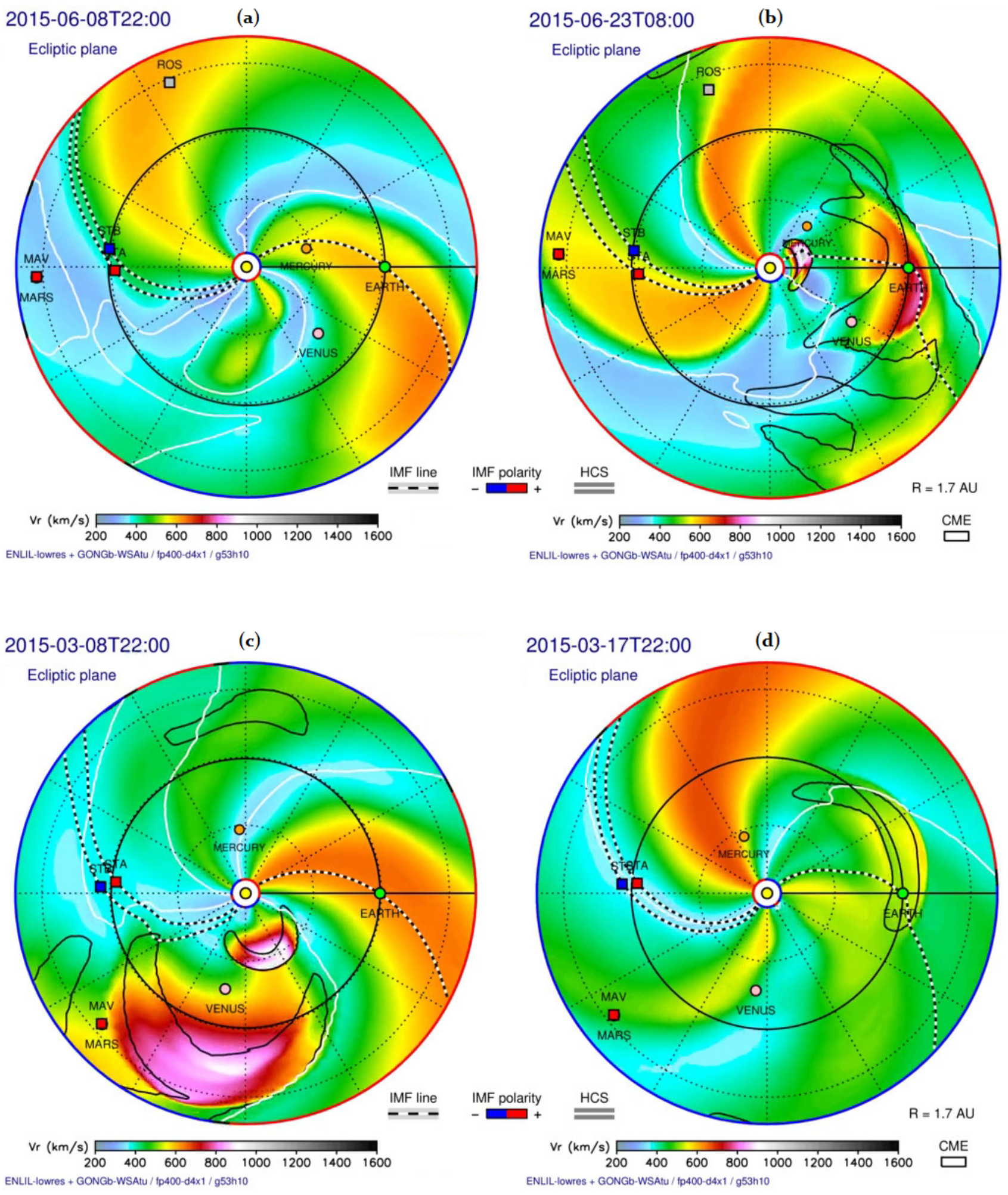}
\caption{WSA-ENLIL simulations of solar wind during (a) 8 June 2015, (b) 23 June 2015, (c) 8 March 
2015, and (d) 17 March 2015. }
\end{figure}

Figures (6a-6d) shows the WSA-ENLIL simulations of solar wind conditions during 
the CIR event on 8 June 2015 and 23 June 
2015 (Figures 6a-6b), and CME events on 8 March 2015 and 17 March 2015 (Figures 6c-6d).\\

\begin{figure}[h]
\centering
\includegraphics[width=1.05\linewidth]{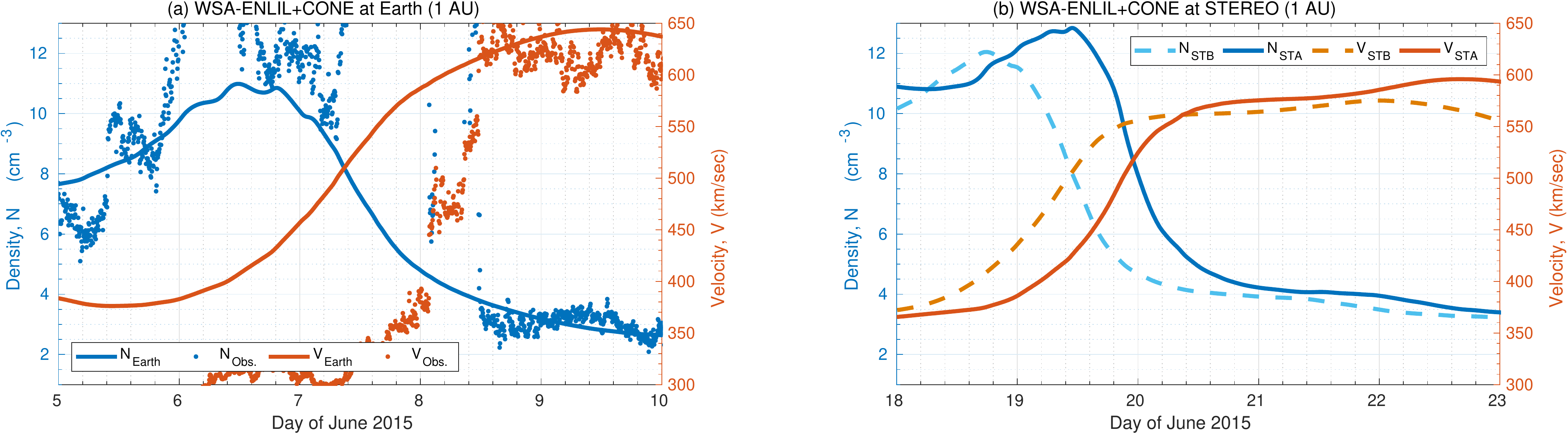}
\caption{WSA-ENLIL+Cone simulations of solar wind density and velocity during (a) 5-9 June 2015 at 
Earth and (b) 18-22 June 2015 at STEREO-A and STEREO-B.}
\end{figure}

Figure 7 shows the WSA-ENLIL+Cone simulations of solar wind density and velocity during 5-9 
June 2015 at Earth (Figure 7a) and 18-22 June 2015 at STEREO-A and STEREO-B (Figure 7b). Figure 
7(a) shows the comparison of ACE observations with WSA-ENLIL+Cone simulation, which are showing a 
gross agreement. We can see that the time evolution of solar wind velocity and density at Earth 
and STEREOs shows an overall similarity, which confirms that the corotating stream was relatively 
steady in time, that is, time stationary in the corotating frame even after half a solar rotation 
\citep{Fisk1980, Reames1997}. It maybe noted here that the WSA-ENLIL+Cone simulations are used only 
to get a general picture of what these events might have looked like on a larger scale at 1 and 1.5 
AU. Though the WSA-ENLIL+Cone can provide short-term all-inclusive forecasts which are closer to 
the observations, the long-term forecasts may be imprecise in details \citep{Lentz2018, 
Falkenberg2011}. However, the predictions are reasonably accurate within 1 AU \citep{Mays2015}; and 
therefore the simulations at STEREO locations (where observations are not available) are expected 
to be reasonable.


\newpage

\acknowledgments
\section*{acknowledgments}
The work is supported by the Indian Space Research Organisation (ISRO). The MAVEN data used in this 
work are taken from the Planetary Data System (\url{https://pds.nasa.gov/}). We gratefully 
acknowledge the MAVEN team for the data. The ACE/EPAM data are  taken from the ACE Science Center 
(\url{http://www.srl.caltech.edu/ACE/ASC/level2/}). The solar wind velocity and IMF near 1 AU as 
well as the GOES particle flux data are obtained from the SPDF OMNIWeb data center 
(\url{https://omniweb.gsfc.nasa.gov/}). We thank the staff of the ACE Science Center for providing 
the ACE  data and NASA/GSFC OMNIWeb team for the interplanetary and GOES data. The WSA-ENLIL+Cone 
simulations are taken from ENLIL Solar Wind 
Prediction (\url{http://helioweather.net/}, \url{https://iswa.ccmc.gsfc.nasa.gov/}). The 
authors thank K. Kishore Kumar, Space Physics Laboratory, VSSC for the useful 
discussions on spectral analysis. The 
authors also thank K. Sankarasubramanian, U R Rao Satellite Centre for the useful discussions on 
SDO solar images. C. Krishnaprasad acknowledges 
the financial assistance provided by ISRO through a research fellowship.

\bibliographystyle{aasjournal}

\end{document}